\documentclass[pre,showpacs,twocolumn,superscriptaddress,letter]{revtex4}
\usepackage{epsfig}

\usepackage{graphicx}
\newcommand{\be}{\begin{equation}} \newcommand{\ee}{\end{equation}}
\newcommand{\bea}{\begin{eqnarray}} \newcommand{\eea}{\end{eqnarray}}
\newcommand{\iint}{\int\!\!\!\!\int}

\begin{document} \title{Lower Bounds on Mutual Information}

\author{David V. Foster} \affiliation{Complexity Science Group, University of Calgary, Calgary, Canada} 
\author{Peter Grassberger} \affiliation{Complexity Science Group, University of Calgary, Calgary, Canada} 
\affiliation{John von Neumann Inst. f\"ur Computing, FZ J\"ulich, D-52425 J\"ulich, Germany}

\begin{abstract} 
  We correct claims about lower bounds on mutual information (MI) between real-valued random 
  variables made in A. Kraskov {\it et al.}, Phys. Rev. E {\bf 69}, 066138 (2004). We 
  show that non-trivial lower bounds on MI in terms of linear correlations depend on 
  the marginal (single variable) distributions. This is so in spite of the invariance of MI under 
  reparametrizations, because linear correlations are not invariant under them. The simplest
  bounds are obtained for Gaussians, but the most interesting ones for practical purposes are obtained  
  for uniform marginal distributions. The latter can be enforced in general by using the ranks
  of the individual variables instead of their actual values, in which case one obtains bounds 
  on MI in terms of Spearman correlation coefficients. We show with gene expression data that these 
  bounds are in general non-trivial, and the degree of their (non-)saturation yields valuable
  insight.
\end{abstract} 

\maketitle

Mutual information \cite{Cover} between two objects is the difference between the combined 
lengths of their individual descriptions and the length of a joint description, all descriptions being 
``optimal", i.e. lossless and redundancy-free. In the framework of algorithmic information theory 
\cite{Li-Vitanyi}, this is taken literally, i.e. the ``objects" are sequences of letters of 
some alphabet, and ``description" means a compression of the sequence on some specified but otherwise 
arbitrary universal Turing machine. In the framework of Shannon theory, in contrast, we 
deal 
with random variables, and ``description length" is to be understood
as the minimal {\it average} information needed to specify their realizations, given the probability 
distributions. 

In the following we shall only use the Shannon framework, but we shall not forget entirely about 
individual objects. When confronted with them, we make some (explicit or implicit) estimate about the 
probability distribution (assuming that the observed objects are in some sense ``typical"); computing
their MI is actually a problem of statistical inference.

More precisely, consider two random variables $X$ and $Y$ with realizations $x,y$ and probability densities 
$p_X(x)$ and $p_Y(y)$. For simplicity we shall assume that $x$ and $y$ are both scalars taken either from
a finite interval or from the interval $[-\infty,\infty]$. In both cases $p_X$ and $p_y$ are normalized to 
1. The joint distribution is $p(x,y)$. The MI is then defined as
\be
   I(X:Y) = \int dx dy\; p(x,y) \log {p(x,y)\over p_X(x)p_Y(y)},
\ee
where the base of the logarithm specifies the units in which information is measured. Bits correspond to 
logarithm base 2.

From this one sees that $I$ is symmetrical, $I(X:Y)=I(Y:X)$, and positive definite: $I(X:Y)=0$ iff $X$ and 
$Y$ are strictly independent. Thus $I(X:Y)$ is a universal measure of dependency, being non-zero whenever
$X$ and $Y$ have anything in common. This can also be seen in the following way: the (differential) 
entropy $H(X)=-\int dx \;p_X(x) \log p_X(x)$ is the (negative) average log-likelihood of $x$, and 
\be
    I(X:Y) = H(X)-H(X|Y)
\ee
is the logarithm of the ratio between the unconditioned likelihood of $x$ and the posterior likelihood
conditioned on the value $y$ of $Y$.

For the differential entropy, there is a well known upper bound in terms of the variance: $H(X)$
is maximal for a Gaussian with the same variance as the data \cite{Cover}. Indeed, this is true also for 
multivariate distributions. In the 
appendix of \cite{Kraskov}, a formal proof based on Lagrangian multipliers was given that 
analogous bounds hold also for the MI. According to \cite{Kraskov}, a given covariance matrix
implies a lower bound on the MI. Unfortunately, this proof is wrong, and the claim made in \cite{Kraskov} is 
incorrect. The error in \cite{Kraskov} was subtle: The unique solution of the Lagrangian variational 
problem was given correctly, but the fact was missed that this solution is in general a saddle point, 
the correct bound being an infimum which is not reached by any actual distribution (at least not by a 
distribution in the class admitted in the variational problem).

Indeed, it is easily seen that the MI can be arbitrarily small for any value of the correlation. Assume 
that the joint distribution is a sum of a delta peak with weight $1-\epsilon$ centered at $(x,y)=(1,1)$
and a 2-d Gaussian with weight $\epsilon$ centered at the origin,
\be
   p(x,y) = (1-\epsilon)\delta(x-1)\delta(y-1) + {\epsilon\over 2\pi\sigma^2} e^{-{x^2+y^2\over 2\sigma^2}}.
\ee
Then the correlation between $X$ and $Y$ varies between zero and one as the width $\sigma$ shrinks to zero,
for any fixed $\epsilon>0$. But the MI is bounded for all $\sigma$ by $I(X:Y) \leq -\epsilon \log\epsilon- 
(1-\epsilon)\log(1-\epsilon)$, which tends to zero as $\epsilon\to 0$. Thus the MI can be arbitrarily close 
to zero, even when the correlation is arbitrarily close to 1 -- although this is unlikely to appear 
in real applications, except for outliers.

It is the purpose of the present paper to present correct bounds replacing those given in \cite{Kraskov}.
As we shall see, to obtain non-trivial bounds for the MI, one needs both the covariance matrix
and the marginal distributions. But the latter can be chosen arbitrarily to 
a large amount, since $I(X:Y)$ as defined in Eq.~(1) is invariant under homeomorphism. Let $\phi(x)$ be 
a continuous and monotonic function, such that its inverse $\phi^{-1}(x)$ is also continuous and monotonic,
and let $X'$ be a random variable with realization $x'=\phi(x)$ if $X$ has realization $x$. Then
\be
   p_X(x) = \left|{d\phi(x)\over dx}\right| p_{X'}(x'),
\ee
and $I(X:Y) = I(X':Y)$. By symmetry, the same holds for homeomorphisms of $Y$. 

This leads to the following strategy for obtaining bounds on $I(X:Y)$: One first transforms $X$ and $Y$
independently so that they have a given distribution, e.g. a Gaussian or a uniform distribution. Notice
that the first and second moments in general will change during such a transformation. After that is done,
one applies the bound suitable for the chosen marginal distributions.

The case of Gaussian marginal distributions is the simplest to treat theoretically. In that case the arguments 
given in the appendix of \cite{Kraskov} apply, and the MI is bounded from below by the MI of a joint
Gaussian with the observed first \& second moments. But this is not the most practical choice, because
it is non-trivial to transform any empirical distribution into a Gaussian.

For practical purposes much more suitable is transformation to uniform distributions over finite intervals,
say $x' \in [-1,1]$ and $y'\in [-1,1]$. This transformation, which also leads usually to improved MI 
estimates, is de facto achieved by using for $x'$ and $y'$ their normalized
ranks. Assume that the empirical data consist of $N$ pairs $(x_i,y_i),\;i=1,\ldots N$. Then the rank $r_i$ of 
$x_i$ is defined as the number of values $x_j$ which are less than or equal to $x_i$ (here we assume that all 
$x_i$ are different, as would be true with probability 1 if $X$ is drawn from a continuous distribution; if there 
are degeneracies due e.g. to discretization, we remove them by adding small random fluctuations to $x_i$).
Finally, 
\be
   x'_i = 2r_i/N - 1.
\ee
and analogously for $y$. Notice that this does not, strictly speaking, define $X'$, as it defines the 
homeomorphism $\phi$ only at the discrete values $x_i$, but this does not pose a practical problem. 
Furthermore, in the limit $N\to\infty$ the ``empirical $\phi(x)$" tends with probability 1 towards
a true homeomorphism. The linear correlation between the ranks of $x$ and $y$ is by definition the 
Spearman coefficient $S = C_{X'Y'}$ \cite{Spearman}.

To obtain a bound on the MI for given marginal distributions and given first \& second moments, we 
use the Lagrangian method. Without loss of generality we assume that the data are centered, i.e. 
$\langle X\rangle= \langle Y\rangle=0$. We use $p(x,y)$ as independent variables, and 
\bea
   p_X(x) &=& \int dy\; p(x,y),\quad p_Y(y)=\int dx\;p(x,y);\nonumber \\
          C_{XY} &=& \iint dx dy\; xy\; p(x,y)/[\sigma_X\sigma_Y]
\eea
as constraints. The Lagrangian function is 
\bea
   L &=& \iint dx dy\; p(x,y) \log {p(x,y)\over p_X(x)p_Y(y)} \nonumber \\
     &+& \int dx\; \nu_X(x)[p_X(x)-\int dy\; p(x,y)] \nonumber \\
     &+& \int dy\; \nu_Y(y)[p_Y(y)-\int dx\; p(x,y)] \nonumber \\
     &+& \lambda [ \sigma_X\sigma_Y C_{XY} - \iint dx dy\; xy p(x,y)].
\eea
where $\nu_X(x)$, $\nu_Y(y)$, and $\lambda$ are Lagrangian parameters. The variational equations are 
\be
   {\delta L\over \delta p(x,y)} = \log {p(x,y)\over p_X(x)p_Y(y)} + 1 - \nu_X(x)-\nu_Y(y) -\lambda xy =0,
\ee 
which can also be written as
\be
   p(x,y) = f_X(x) f_Y(y) e^{-\lambda (x-y)^2}               \label{pxy}
\ee
with unknown functions $f_X,f_Y$ and unknown $\lambda$, all of which are determined by the constraints.
The Kolmogorov consistency condition for $p_X(x)$, in particular, gives
\be
   {p_X(x)\over f_X(x)} = \int dy\; f_Y(y)e^{-\lambda (x-y)^2}.
\ee

In the following we shall only discuss the two cases of Gaussian and uniform marginals. For Gaussian 
marginals, one finds that $p(x,y)$ is also Gaussian, and thus the results of \cite{Kraskov} are obtained,
\be
   I(X:Y)\geq I_{\rm Gauss}^-(C_{XY}) \equiv -{1\over 2}\log(1-C_{XY}^2).     \label{Gauss}
\ee
For uniform marginals, we indeed do not solve the problem of finding a bound $I_{\rm unif}^-$ on the MI 
for given $S$, but we solve the easier implicit problem of finding both $I_{\rm unif}^-$ and $S$ for given 
$\lambda$. We do this recursively, starting with the zeroth approximation
\be
   f_X^{(0)}(x)= f_Y^{(0)}(y) = 1/2.
\ee
From the $k$-th approximation of $f_X$ and $f_Y$ we obtain the $(k+1)$-st approximations
by means of 
\be
   {1\over f^{(k+1)}_X(x)} = 2\int_{-1}^1 dy\; f^{(k)}_Y(y)e^{-\lambda (x-y)^2},                \label{a}
\ee
\be
   {1\over f^{(k+1)}_Y(y)} = 2\int_{-1}^1 dx\; f^{(k)}_X(x)e^{-\lambda (x-y)^2}.                \label{b}
\ee
When doing this, we observe that $f^{(k)}_X$ and $f^{(k)}_Y$ are even functions for each $k$, 
and that both indeed are equal. We can thus drop the subscripts and write the recursion as
\be
   f^{(k+1)}(x) = \left[2 \int_{-1}^1 dy\; f^{(k)}(y) e^{-\lambda (x-y)^2}\right]^{-1}.
\ee
After convergence, the joint density is obtained as
\be
   p(x,y) \propto \lim_{k\to\infty} f^{(k)}(x) f^{(k)}(x) e^{-\lambda (x-y)^2}.
\ee
Here we have left the normalization open, in order to allow for errors in the numerical
integration which might have accumulated during the recursion. The proportionality constant 
is thus fixed by the normalization condition $\iint p=1$.
Finally, $S$ and the lower bound $I_{\rm unif}^-(S)$ on $I(X:Y)$ are obtained by using Eq.~(1) and
\be
   S = 3 \iint_{-1}^1 dxdy\; xy\; p(x,y) e^{-\lambda (x-y)^2}.               \label{c}
\ee

\begin{table}
\begin{center}
\begin{tabular}{|c|c|c|} \hline
  $\;\;\quad\lambda \qquad $ & $\;\;\quad  S \qquad $  &  $\;\;\quad  I^-_{\rm unif} \qquad $\\ \hline 
   0.00 &  0.0000 &  0.0000 \\
   0.25 &  0.0829 &  0.0034 \\
   0.50 &  0.1633 &  0.0135 \\
   0.75 &  0.2390 &  0.0292 \\
   1.00 &  0.3086 &  0.0495 \\
   1.25 &  0.3713 &  0.0729 \\
   1.50 &  0.4270 &  0.0984 \\
   2.00 &  0.5189 &  0.1517 \\
   2.50 &  0.5897 &  0.2040 \\
   3.00 &  0.6428 &  0.2531 \\
   4.00 &  0.7177 &  0.3396 \\
   5.00 &  0.7666 &  0.4123 \\
   6.00 &  0.8007 &  0.4746 \\
   7.00 &  0.8260 &  0.5292 \\
   8.00 &  0.8455 &  0.5777 \\
   9.00 &  0.8610 &  0.6215 \\
  10.00 &  0.8736 &  0.6614 \\
  11.50 &  0.8887 &  0.7156 \\
  13.00 &  0.9005 &  0.7636 \\
  15.00 &  0.9128 &  0.8208 \\
  17.00 &  0.9224 &  0.8717 \\
  20.00 &  0.9333 &  0.9389 \\
  23.00 &  0.9415 &  0.9975 \\
  27.00 &  0.9498 &  1.0657 \\
  32.00 &  0.9572 &  1.1393 \\
  40.00 &  0.9654 &  1.2366 \\
  50.00 &  0.9721 &  1.3357 \\ \hline
\end{tabular}
\caption{Spearman coefficient and lower bound on the MI (in natural units).}
\end{center}
\end{table}

Numerical results for several values of $\lambda$, obtained by using Gaussian 
quadrature for the integrals, are given in Table 1. Except for values of $S$ close to 
$\pm 1$, $I_{\rm unif}^-(S)$ is well approximated by
\be
   I^-_{\rm unif}(S) \approx -{1\over 2} (1-0.122\;S^2+0.053\;S^{12})\log(1-S^2).     \label{approx}
\ee
The two bounds for Gaussians [Eq.~(\ref{Gauss})] and for uniform distributions 
[Eq.~(\ref{approx})] are shown in Fig.~1.

\begin{figure}
  \centering
  \includegraphics[width=4.6cm,angle=270]{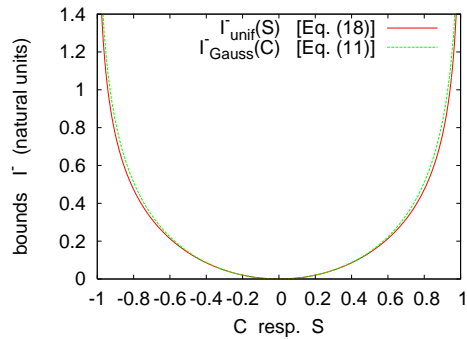}
  \caption{(color online) Lower bounds of MI in terms of the Spearman correlation coefficient
  (continuous line, red) and in terms of the Pearson correlation coefficient in case of Gaussian 
  marginals (dashed, green). For both curves, the MI is measured in natural units.}
  \label{fig:bound}
\end{figure}

\begin{figure}
  \centering
  \includegraphics[width=5.7cm,angle=270]{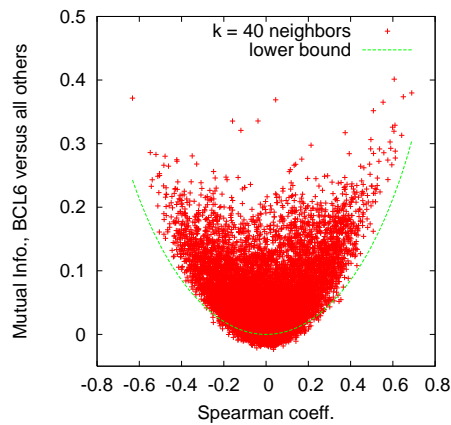}
  \caption{(color online) Mutual informations between gene BCL6 and all the other 12599 genes
  as measured in the microarray gene expression experiment of \cite{Basso}. Values of the 
  MI were estimated by means of $k$-nearest neighbors with $k=40$. The green line is the 
  lower bound discussed in this paper.}
  \label{fig:BCL6}
\end{figure}

\begin{figure}
  \centering
  \includegraphics[width=5.5cm,angle=270]{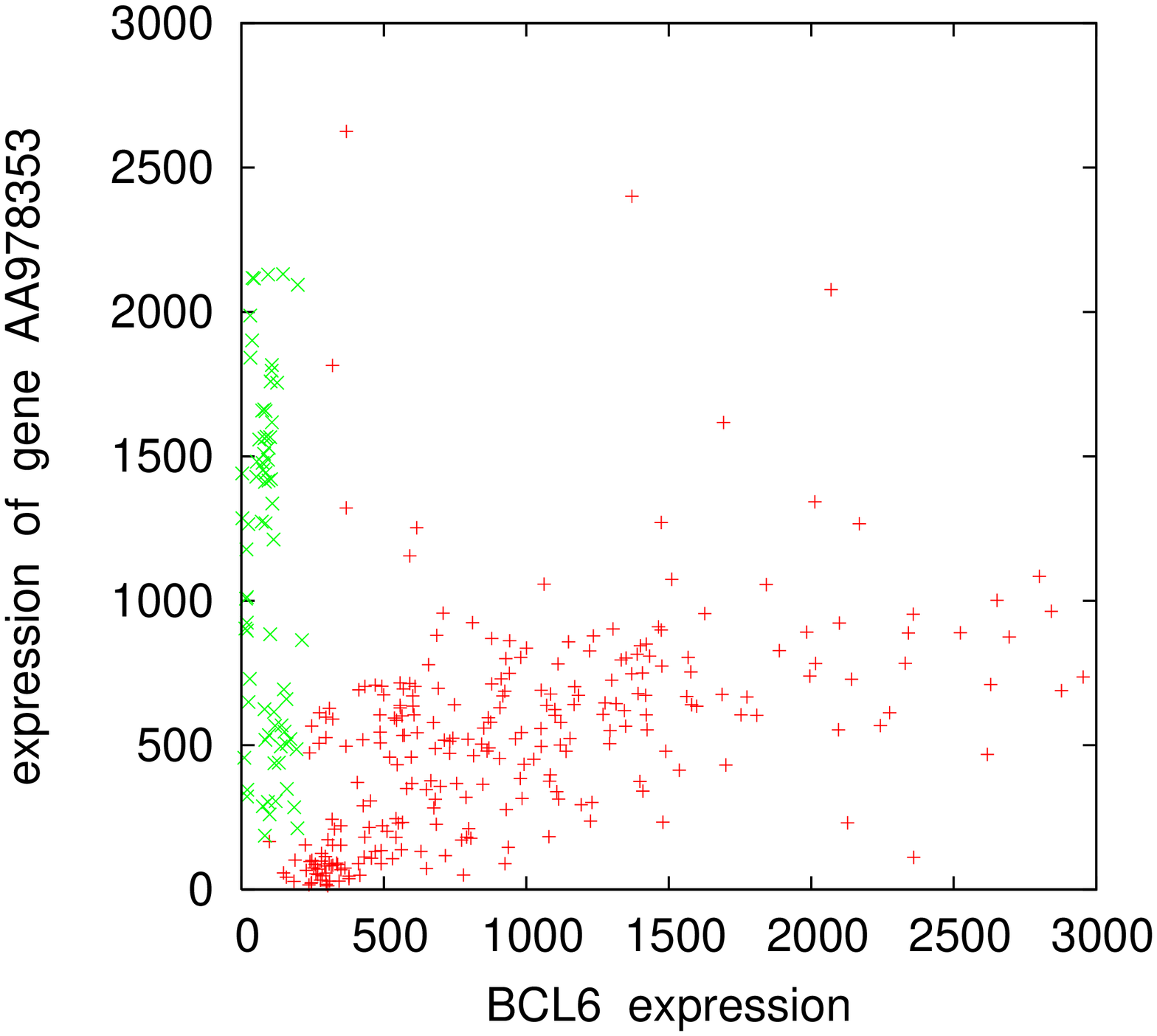}
  \includegraphics[width=5.5cm,angle=270]{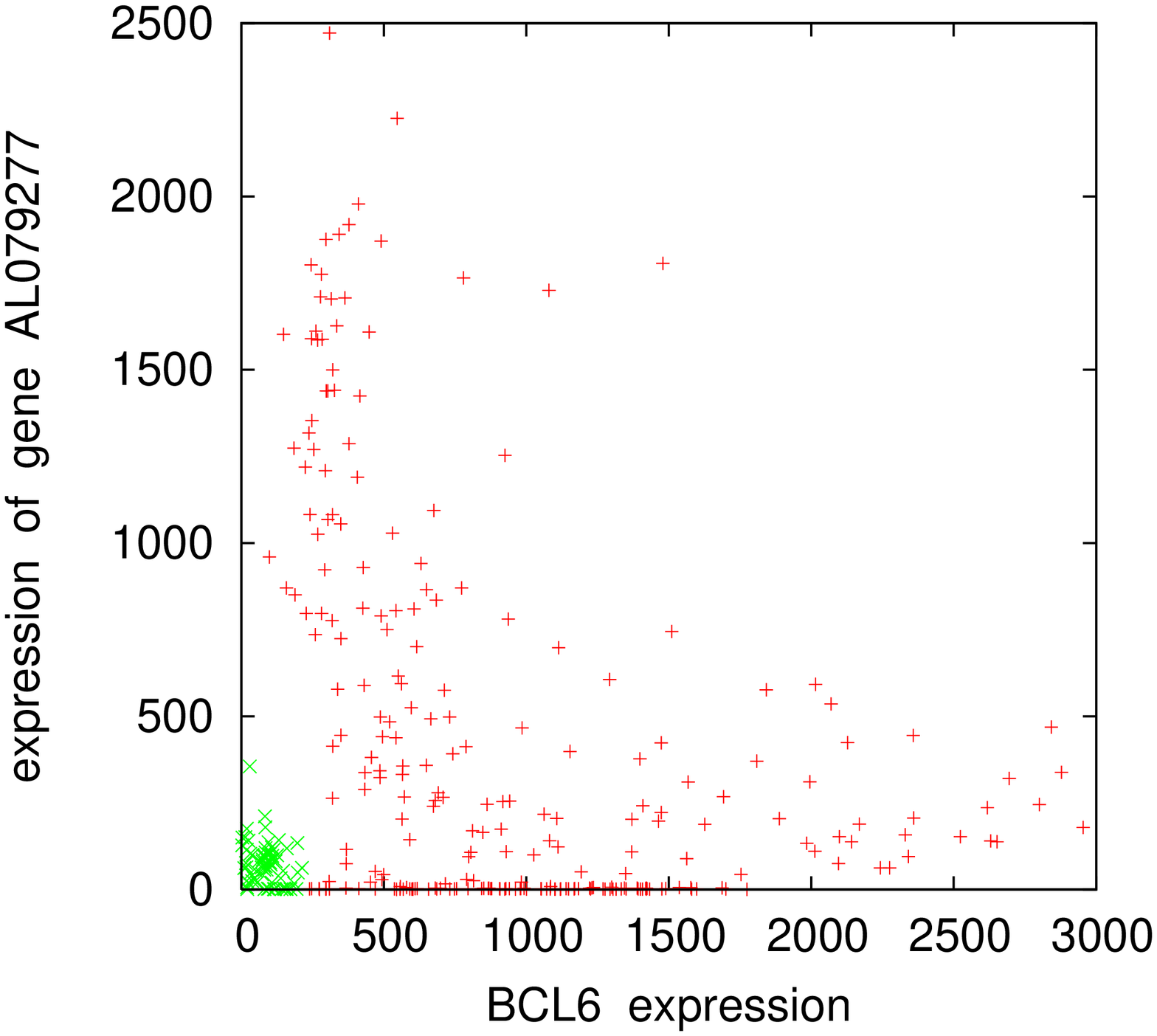}
  \caption{(color online) Each panel shows the gene expression intensities (arbitrary units) 
  of two genes, one of which is BCL6 (x-axis). The other gene (y-axis) was chosen such as 
  to have very large MI with BCL6, but very small Spearman coefficient (the uppermost two 
  points in Fig.~2 with $|S|<0.1$). The color coding (green for BCL6 expression $<200$ and
  $AL079277$ expression $<500$, red otherwise) is such that the 
  same cell conditions have in both panels the same color. It suggests that the observed
  nonlinear correlations are related to the existence of two cell populations with very
  different properties. The two genes correspond to accession numbers $AA978353$ (top)
  and and $AL079277$ (bottom).}
  \label{fig:igenes-BCL6}
\end{figure}

As an application we show in Fig.~2 gene expression data obtained from human B lymphocyte
cells \cite{Basso}. In that experiment, the expressions of 12600 different gene loci were 
measured in 336 different conditions, with special interest in tumor cells. For each pair 
of genes the data can thus be represented as 336 points in a two-dimensional plane. 
Spearman coefficients were obtained by ranking both coordinates (after disambiguating 
degeneracies by adding low level noise as explained above). Mutual informations were 
estimated using the $k$-nearest neighbor method of \cite{Kraskov} with $k=40$. Although 
this was done for all $12600 \times 12599/2$ pairs, only results for the 12599 pairs 
involving the important cancer gene BCL6 are shown in Fig.~2. We can make the following 
observations:

\begin{itemize}
\item The bound is respected by most pairs, and it forms roughly a lower 
envelope for the distribution.
\item There are several pairs for which the bound is violated, mostly for small values of $S$.
This reflects the fact that the MI estimator is not perfect. Indeed, no MI estimator 
can be perfect. Most estimators are chosen such that they never produce negative MI, which 
is achieved by tolerating a positive bias. The estimator of \cite{Kraskov} was constructed
such that the bias is minimized, at the cost of obtaining occasionally negative 
values due to statistical fluctuations.
\item For most pairs the bound is not saturated, showing that there are important non-linear 
dependencies between these pairs. As an illustration for the latter we take the two 
points with $|S|<0.1$ and $I>0.3$ and plot their gene expression vectors in Fig.~3. They 
show the co-expression of BCL6 with the genes with GenBank accession numbers $AA978353$ (top) 
and $AL079277$ (bottom). In
both panels of Fig.~3 we see very strong dependencies which cannot be approximated
by linear correlations. Neither of these two genes is known to be related to BCL6, maybe 
because such relations were overlooked because of the small linear correlations. The 
data suggest the presence of (at least) two different 
sub-populations of cells, marked in Fig.~3 by different colors. In the sub-population in 
which BCL6 is strongly expressed (red points in Fig.~3) there are also significant linear 
correlations.
\end{itemize} 

In summary, we have derived lower bounds on the MI between real-valued variables in terms of 
linear correlation coefficients. We have seen that such bounds are not independent of the 
marginal distribution, in contrast to the claims made in the appendix of \cite{Kraskov}. 
But one can use the homeomorphism invariance of the MI to transform the variables to
new variables with uniform distribution, in which case the linear correlation coefficient 
becomes equal to the Spearman coefficient $S$. At least in one specific and 
scientifically relevant example, the resulting bound of the MI in terms of $S$ was 
found to be numerically non-trivial. In particular, large discrepancies between the 
bound and the actual values gave hints to specific structures in the data which then 
could be investigated in more detail. The bound can also be useful in 
testing MI estimators. Usually, an estimator is deemed unacceptable if it violates the 
bound $I(X:Y) \geq 0$. But it would be equally unacceptable, if it violates the 
stronger bound $I(X:Y) \geq I^-$. 

Finally, our results also answer the question of how linear correlations 
change under reparametrizations. There is no reason to expect a universal exact answer, but 
approximately they should change such that the numerical values of the bounds $I^-$ stay
the same.

We thank Andrea Califano for providing us the data of Ref.~\cite{Basso}, and Alexander 
Kraskov and Maya Paczuski for discussions.

\end{document}